\documentclass[twocolumn,preprintnumbers,amsmath,amssymb]{revtex4}

\usepackage{graphicx}
\usepackage{dcolumn}
\usepackage{bm}
\usepackage{rotating}

\usepackage{epstopdf}
\usepackage{bbm}
\usepackage{verbatim}
\usepackage{txfonts}
\usepackage[cspex,bbgreekl]{mathbbol}

\usepackage{color}
\definecolor{lightblue}{rgb}{0.2,0.2,0.7}
\definecolor{darkblue}{rgb}{0,0.25,0.5}
\definecolor{redbrown}{rgb}{0.875,0.25,0.125}
\definecolor{darkgreen}{rgb}{0,0.5,0}

\newcommand{\bra}[1]{\ensuremath{\langle #1 \vert}}
\newcommand{\ket}[1]{\ensuremath{\vert #1  \rangle}}
\newcommand{\braket}[2]{\ensuremath{\langle  #1 \vert #2  \rangle}}
\renewcommand{\b}[1]{\ensuremath{\mathbf{#1}}}
\renewcommand{\H}{\ensuremath{\text{H}}}

\newcommand{\tr}{\ensuremath{\text{tr}}}
\newcommand{\lr}{\ensuremath{\text{lr}}}
\newcommand{\sr}{\ensuremath{\text{sr}}}

\newcommand{\dRPA}{\ensuremath{\text{dRPA}}}
\newcommand{\RPAx}{\ensuremath{\text{RPAx}}}

\newcommand{\RPAxII}{\ensuremath{\text{RPAx-II}}}

\newcommand{\RPAxSOo}{\ensuremath{\text{RPAx-SO1}}}
\newcommand{\RPAxSOt}{\ensuremath{\text{RPAx-SO2}}}
\newcommand{\SOSEX}{\ensuremath{\text{SOSEX}}}
\newcommand{\CCD}{\ensuremath{\text{CCD}}}

\newcommand{\R}{\ensuremath{\text{R}}}

\newcommand{\ext}{\ensuremath{\text{ext}}}

\newcommand{\T}{\ensuremath{\text{T}}}
\renewcommand{\S}{\ensuremath{\text{S}}}

\newcommand{\ua}{\ensuremath{\uparrow}}
\newcommand{\da}{\ensuremath{\downarrow}}

\begin{document}

\title{Closed-shell ring coupled cluster doubles theory with range separation applied on weak intermolecular interactions}

\author{Julien Toulouse$^{1}$}\email{julien.toulouse@upmc.fr}
\author{Wuming Zhu$^{2}$}
\author{Andreas Savin$^{1}$}
\author{Georg Jansen$^{3}$}
\author{J\'anos G. \'Angy\'an$^{4}$}
\affiliation{
$^1$ Laboratoire de Chimie Th\'eorique, Universit\'e Pierre et Marie Curie and CNRS, 75005 Paris, France\\
$^2$ Department of Physics, Hangzhou Normal University, HangZhou XiaSha, 310036 ZheJiang, China\\
$^3$ Fakult\"at f\"ur Chemie, Universit\"at Duisburg-Essen, 45117 Essen, Germany\\
$^4$ CRM2, Institut Jean Barriol, Nancy University and CNRS, 54506 Vandoeuvre-l\`{e}s-Nancy, France
}


\date{\today}
\begin{abstract}
We explore different variants of the random phase approximation (RPA) to the correlation energy derived from closed-shell ring-diagram approximations to coupled cluster doubles theory. We implement these variants in range-separated density-functional theory, i.e. by combining the long-range random phase approximations with short-range density-functional approximations. We perform tests on the rare-gas dimers He$_2$, Ne$_2$, and Ar$_2$, and on the weakly interacting molecular complexes of the S22 set of Jure\v cka {\it et al.} [Phys. Chem. Chem. Phys. {\bf 8}, 1985 (2006͒)]. The two best variants correspond to the ones originally proposed by Szabo and Ostlund [J. Chem. Phys. {\bf 67}, 4351 (1977)]. With range separation, they reach mean absolute errors on the equilibrium interaction energies of the S22 set of about 0.4 kcal/mol, corresponding to mean absolute percentage errors of about 4\%, with the aug-cc-pVDZ basis set.
\end{abstract}

\maketitle

\section{Introduction}

In the last decade, there has been a revived interest in the random phase approximation (RPA) and other related approximations for calculating the electron correlation energy of atomic, molecular and solid-state systems~\cite{YanPerKur-PRB-00,Fur-PRB-01,AryMiyTer-PRL-02,MiyAryKotSchUsuTer-PRB-02,FucGon-PRB-02,NiqGon-PRB-04,FucNiqGonBur-JCP-05,FurVoo-JCP-05,DahLeeBar-PRA-06,MarGarRub-PRL-06,JiaEng-JCP-07,HarKre-PRB-08,Fur-JCP-08,ScuHenSor-JCP-08,TouGerJanSavAng-PRL-09,JanHenScu-JCP-09,JanHenScu-JCP-09b,JanScu-JCP-09,RenRinSch-PRB-09,LuLiRocGal-PRL-09,HarKre-PRL-09,NguGir-PRB-09,GruMarHarSchKre-JCP-09,RuzPerCso-JCTC-10,NguGal-JCP-10,HelBar-JCP-10,HesGor-MP-10,PaiJanHenScuGruKre-JCP-10,HarSchKre-PRB-10,Ism-PRB-10,ZhuTouSavAng-JCP-10,EshYarFur-JCP-10,TouZhuAngSav-PRA-10,JanLiuAng-JCP-10,LuNguGal-JCP-10,HesGor-PRL-11,RuzPerCso-JCP-11,RenTkaRinSch-PRL-11,LotBar-JCP-11,KloTeaCorPedHel-CPL-11,Hes-JCP-11}. One particularly appealing feature of RPA is its correct description of dispersion forces at large separation~\cite{DobMcLRubWanGouLeDin-AJC-01,DobWanDinMclLe-IJQC-05,Dob-JCTN-09}. However, RPA is a poor approximation to short-range correlations~\cite{YanPerKur-PRB-00}, and, in a Gaussian localized basis, RPA calculations have a slow convergence with respect to the basis size~\cite{Fur-PRB-01}. A promising strategy is thus to combine a long-range RPA-type approximation with a short-range density-functional approximation~\cite{TouGerJanSavAng-PRL-09,JanHenScu-JCP-09,JanHenScu-JCP-09b,PaiJanHenScuGruKre-JCP-10,ZhuTouSavAng-JCP-10,TouZhuAngSav-PRA-10}, hence avoiding the inaccurate description and slow basis-set convergence of short-range correlations in RPA.

Among the different formulations of RPA, the one based on a ring-diagram approximation to coupled cluster doubles (CCD) theory~\cite{San-PL-65,Fre-PRB-77,MosJezSza-IJQC-93,ScuHenSor-JCP-08} is particularly attractive since it avoids the numerical integration over the adiabatic connection, and in principle is amenable to a fast algorithm~\cite{ScuHenSor-JCP-08}. However, due to the fact that the ring approximation breaks the antisymmetry property of the coupled-cluster amplitudes, several non-equivalent variants of ring CCD can be constructed, especially when the exchange terms are included. In this paper, we explore these various ring CCD variants for closed-shell systems, and show that some of them correspond to the RPA correlation energy expressions originally proposed by Szabo and Ostlund~\cite{SzaOst-IJQC-77,SzaOst-JCP-77}. We apply these closed-shell ring CCD variants in the context of range-separated density-functional theory, and test them on rare-gas dimers and on the weakly interacting molecular complexes of the S22 set of Jure\v cka {\it et al.}~\cite{JurSpoCerHob-PCCP-06}.

\section{Theory}

We first show how to rigorously combine a long-range CCD calculation with a short-range density functional (for details on range-separated density-functional theory, see e.g. Refs.~\onlinecite{TouColSav-PRA-04,AngGerSavTou-PRA-05,TouZhuAngSav-PRA-10}). We start from a self-consistent range-separated hybrid (RSH) calculation~\cite{AngGerSavTou-PRA-05} 
\begin{equation}
E_{\text{RSH}} = \min_{\Phi} \{ \langle \Phi | \hat{T} + \hat{V}_{\ext} + \hat{W}^{\lr}_{ee} | \Phi \rangle + E^{\sr}_{\H xc}[n_{\Phi}] \}, 
\end{equation}
where $\hat{T}$ is the kinetic energy operator, $\hat{V}_{\ext}$ is the external potential operator (e.g., nuclei-electron interaction), $\hat{W}_{ee}^{\lr}$ is a long-range electron-electron interaction operator, $E_{\H xc}^{\sr}[n]$ is the associated short-range Hartree-exchange-correlation density functional, and $\Phi$ is a single-determinant wave function with density $n_{\Phi}$. The long-range interaction is constructed with the error function, $w^{\lr}_{ee}(r) = \text{erf}(\mu r)/r$, where $\mu$ is a parameter whose inverse gives the range of the separation. The minimizing RSH single-determinant wave function is denoted by $\Phi_0$ and its associated (approximate) density by $n_0$. In principle, the exact ground-state energy can be obtained from the RSH energy by adding the long-range correlation energy $E_c^{\lr}$
\begin{equation}
E = E_{\text{RSH}} + E_c^{\lr}. 
\end{equation}
Several formally exact expressions can be derived for $E_c^{\lr}$. The one that is most convenient for applying coupled-cluster theory is
\begin{eqnarray}
E_c^{\lr} &=& \bra{\Psi^\lr} \hat{H}^\lr[n] \ket{\Psi^\lr} - \bra{\Phi_0} \hat{H}^\lr[n] \ket{\Phi_0} 
\nonumber\\
&&+ \Delta E^{\sr}_{\H xc} - \int v^{\sr}_{\H xc}[n](\b{r}) \, \Delta n(\b{r}) d\b{r},
\label{Eclr}
\end{eqnarray}
where $\Psi^\lr$ is the ground-state wave function of the long-range interacting Hamiltonian $\hat{H}^\lr[n]=\hat{T} + \hat{W}^{\lr}_{ee} + \hat{V}_{\ext} + \hat{V}^{\sr}_{\H xc}[n]$ with the short-range Hartree-exchange-correlation potential operator $\hat{V}^{\sr}_{\H xc}[n]=\int v^{\sr}_{\H xc}[n](\b{r}) \, \hat{n}(\b{r}) d\b{r}$ written with the density operator $\hat{n}(\b{r})$ and $v^{\sr}_{\H xc}[n](\b{r})=\delta E^{\sr}_{\H xc}[n]/\delta n(\b{r})$. The long-range wave function $\Psi^\lr$ is associated with the exact density $n$. In Eq.~(\ref{Eclr}), the last two terms are the variation of the energy functional, $\Delta E^{\sr}_{\H xc} = E^{\sr}_{\H xc}[n] - E^{\sr}_{\H xc}[n_0]$, and the variation of the associated potential expectation value due to the variation of the density from the RSH one to the exact one, $\Delta n = n - n_0$. The contribution of these last two terms is expected to be small since it is of second order in $\Delta n$
\begin{eqnarray}
\Delta E^{\sr}_{\H xc} - \int v^{\sr}_{\H xc}[n](\b{r}) \, \Delta n(\b{r}) d\b{r} = \,\,\,\,\,\,\,\,
\nonumber\\
 -\frac{1}{2} \iint \frac{\delta E_{\H xc}^{\sr}[n_0]}{\delta n(\b{r}) \delta n(\b{r}')} \Delta n(\b{r}) \Delta n(\b{r}') d\b{r} d\b{r}' + {\cal O}(\Delta n^3).
\label{DeltaE}
\end{eqnarray}

Using a spin-unrestricted CCD ansatz (see Appendix~\ref{app:CCD} for a review of CCD theory) for the long-range wave function, $\ket{\Psi^\lr_{\CCD}} = \exp \left(\hat{T}_2\right) \ket{\Phi_0}$, where $\hat{T}_2 = (1/4) \sum_{ijab} (t_{ij}^{ab})^\lr \hat{a}_a^\dag \hat{a}_i \hat{a}_b^\dag \hat{a}_j$ is the cluster operator for double excitations written in terms of the long-range amplitudes $(t_{ij}^{ab})^\lr$, and occupied ($i$, $j$) and virtual ($a$, $b$) RSH spin-orbital creation and annihilation operators, we approximate the long-range correlation energy as
\begin{eqnarray}
E_{c,\CCD}^{\lr} &=& \bra{\Phi_0} \hat{H}^\lr[n_0] \ket{\Psi^\lr_{\CCD}} - \bra{\Phi_0} \hat{H}^\lr[n_0] \ket{\Phi_0}.
\label{EclrCCD}
\end{eqnarray}
In Eq.~(\ref{EclrCCD}), the variation of the density has been neglected, $n \approx n_0$ (and thus the contribution of Eq.~(\ref{DeltaE}) vanishes), which seems appropriate if we define the coupled-cluster density as the projected one, $\bra{\Phi_0} \hat{n}(\b{r}) \ket{\Psi^\lr_{\CCD}} = \bra{\Phi_0} \hat{n}(\b{r}) \ket{\Phi_0} = n_0(\b{r})$, which does not vary since the CCD wave function does not contain single excitations. The long-range correlation energy can be calculated as, for real spin orbitals,
\begin{eqnarray}
E_{c,\CCD}^{\lr} = \frac{1}{4} \tr \left[ \b{B}^\lr \b{T}^\lr \right] = \frac{1}{2} \tr \left[ \b{K}^\lr \b{T}^\lr \right],
\label{EcCCDlr}
\end{eqnarray}
where $B_{ia,jb}^\lr=\langle ab ||ij \rangle^\lr$ and $K_{ia,jb}^\lr=\langle ab |ij \rangle^\lr$ are the matrices of antisymmetrized and non-antisymmetrized two-electron integrals with long-range interaction $w^{\lr}_{ee}(r)$, respectively, and $T_{ia,jb}^\lr=(t_{ij}^{ab})^\lr$ is the amplitude matrix. The second equality in Eq.~(\ref{EcCCDlr}) is due to the antisymmetry property of the coupled-cluster amplitudes $T_{ia,jb}^\lr$ with respect to the exchange of the indices $i$ and $j$. These amplitudes can be determined by the usual coupled-cluster equations, replacing the normal Hamiltonian by the long-range one $\hat{H}^\lr[n_0]$, which amounts to using the RSH orbital eigenvalues and the long-range two-electron integrals. The present range-separated CCD method can be seen a special case of the more general range-separated coupled-cluster approach of Goll {\it et al.}~\cite{GolWerSto-PCCP-05} which also includes single excitations and possibly perturbative triples.

We now consider the ring-diagram approximation for closed-shell systems. A number of closed-shell ring CCD variants can be defined. In the ring approximation without exchange terms, the direct RPA (dRPA, also sometimes referred to as RPA or time-dependent Hartree) amplitudes for spin-singlet excitations, ${^1}\b{T}^{\lr}_{\dRPA}$, are obtained by the following Riccati equation~\cite{ScuHenSor-JCP-08}
\begin{equation}
{^1}\b{K}^{\lr} + {^1}\b{L}^{\lr}  \, {^1}\b{T}^{\lr}_{\dRPA} + {^1}\b{T}^{\lr}_{\dRPA} {^1}\b{L}^{\lr} + {^1}\b{T}^{\lr}_{\dRPA} \, {^1}\b{K}^{\lr} \, {^1}\b{T}^{\lr}_{\dRPA} = \b{0},
\label{eq1TdRPA}
\end{equation}
with the spin-adapted matrices ${^1}K_{ia,jb}^\lr=2\langle ab |ij \rangle^\lr$ and ${^1}L^{\lr}_{ia,jb}= \Delta \epsilon_{ia,jb}+{^1}K^{\lr}_{ia,jb}$, where $\Delta \epsilon_{ia,jb}=(\epsilon_a-\epsilon_i) \delta_{ij} \delta_{ab}$ is the matrix of the RSH orbital eigenvalue differences ($i,j$ and $a,b$ refer now to occupied and virtual spatial orbitals, respectively). Contracting the dRPA amplitudes with the non-antisymmetrized two-electron integrals ${^1}\b{K}^{\lr}$ gives the dRPA long-range correlation energy (also referred to as dRPA-I in Ref.~\onlinecite{AngLiuTouJan-JJJ-XX})
\begin{equation}
E_{c,\dRPA}^{\lr} = \frac{1}{2} \tr \left[ {^1}\b{K}^{\lr} \, {^1}\b{T}^{\lr}_{\dRPA} \right].
\label{EcdRPAlr}
\end{equation}
Contracting the dRPA amplitudes with the spin-singlet-adapted antisymmetrized two-electron integrals ${^1}B^{\lr}_{ia,jb}=2\langle ab |ij \rangle^\lr-\langle ab |ji \rangle^\lr$ gives the dRPA+SOSEX (or just SOSEX for short) long-range correlation energy~\cite{GruMarHarSchKre-JCP-09,PaiJanHenScuGruKre-JCP-10}
\begin{equation}
E_{c,\SOSEX}^{\lr} = \frac{1}{2} \tr \left[ {^1}\b{B}^{\lr} \, {^1}\b{T}^{\lr}_{\dRPA} \right].
\label{EcSOSEXlr}
\end{equation}

\begin{figure*}
\includegraphics[scale=0.30,angle=-90]{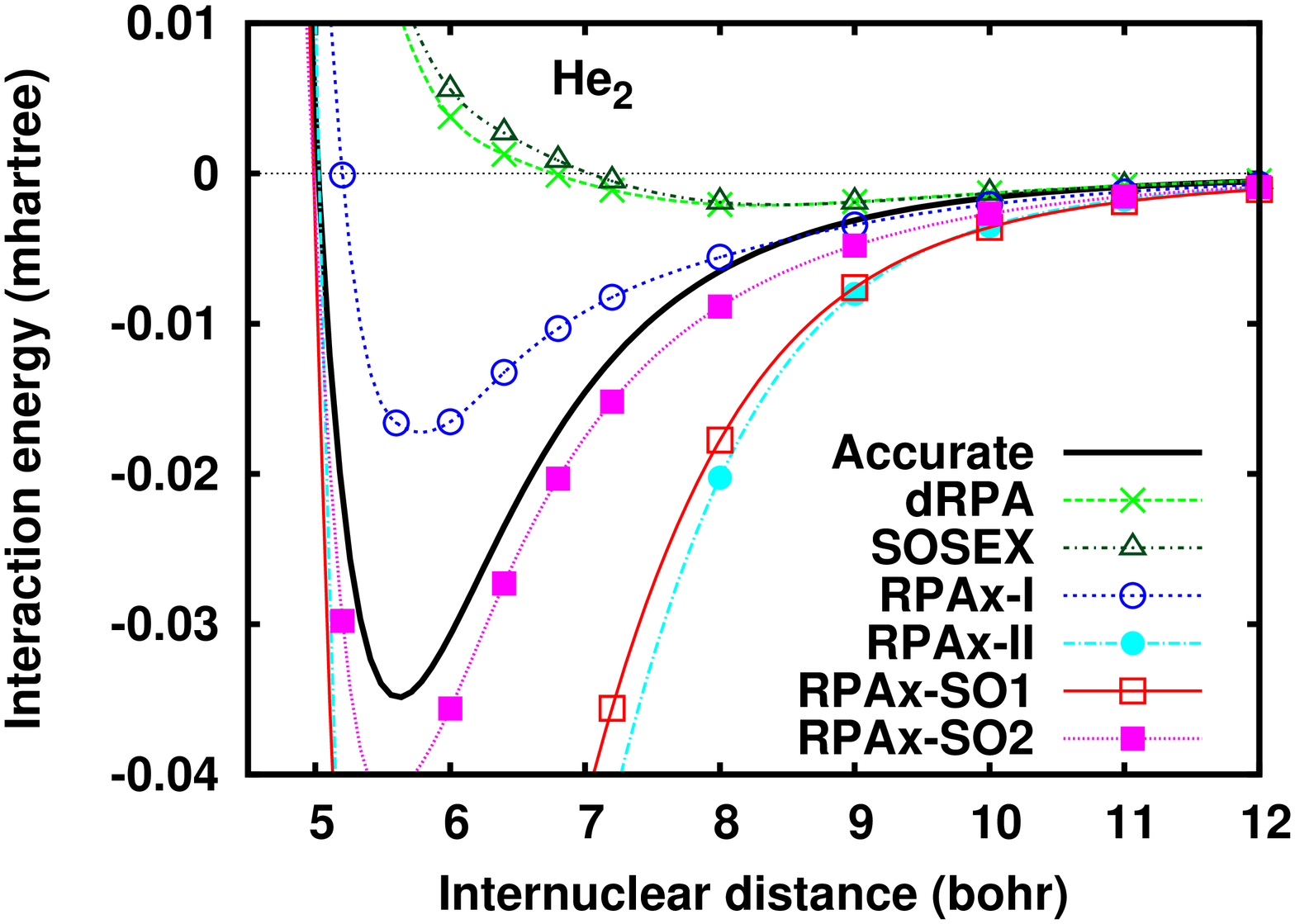}
\includegraphics[scale=0.30,angle=-90]{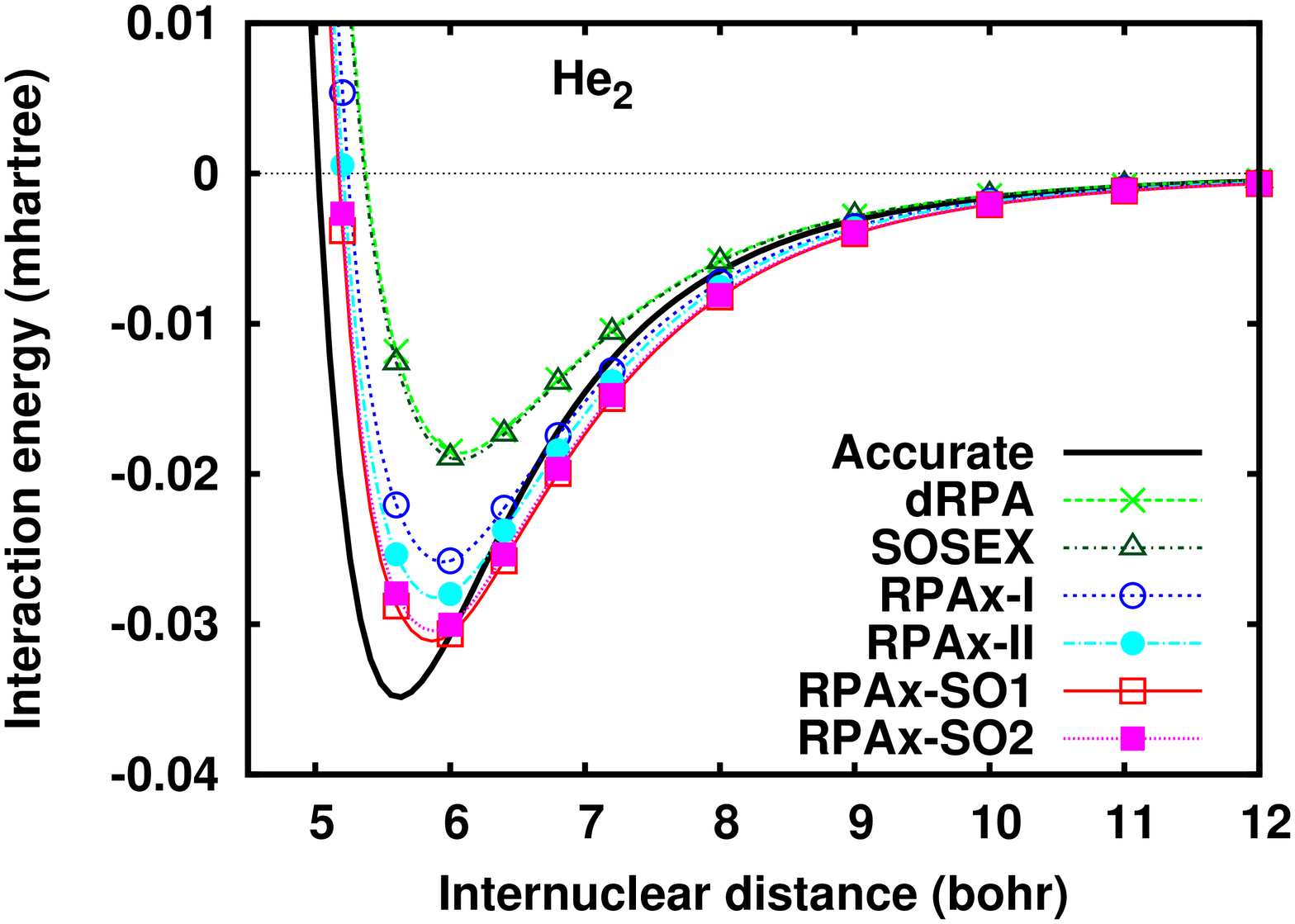}
\includegraphics[scale=0.30,angle=-90]{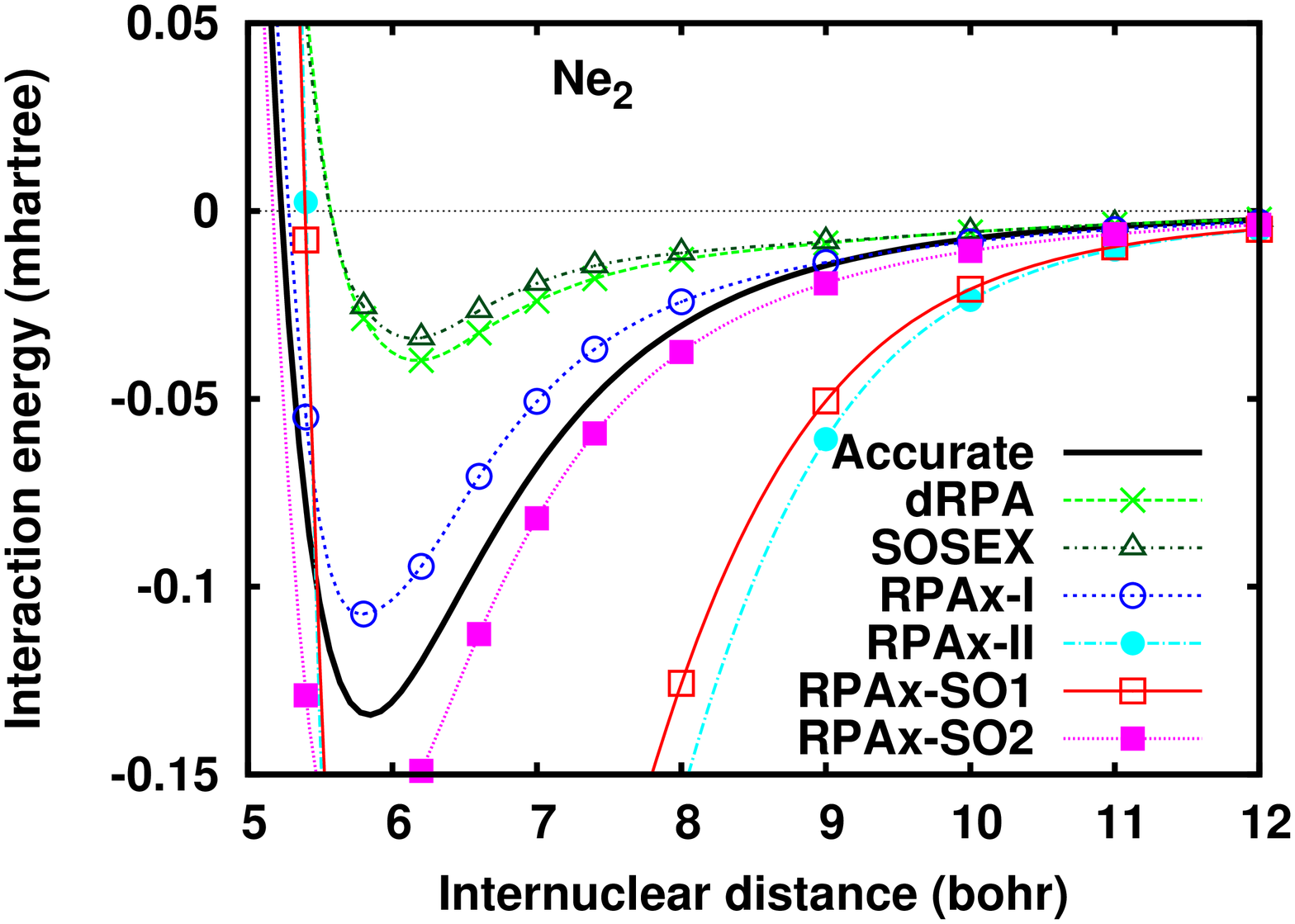}
\includegraphics[scale=0.30,angle=-90]{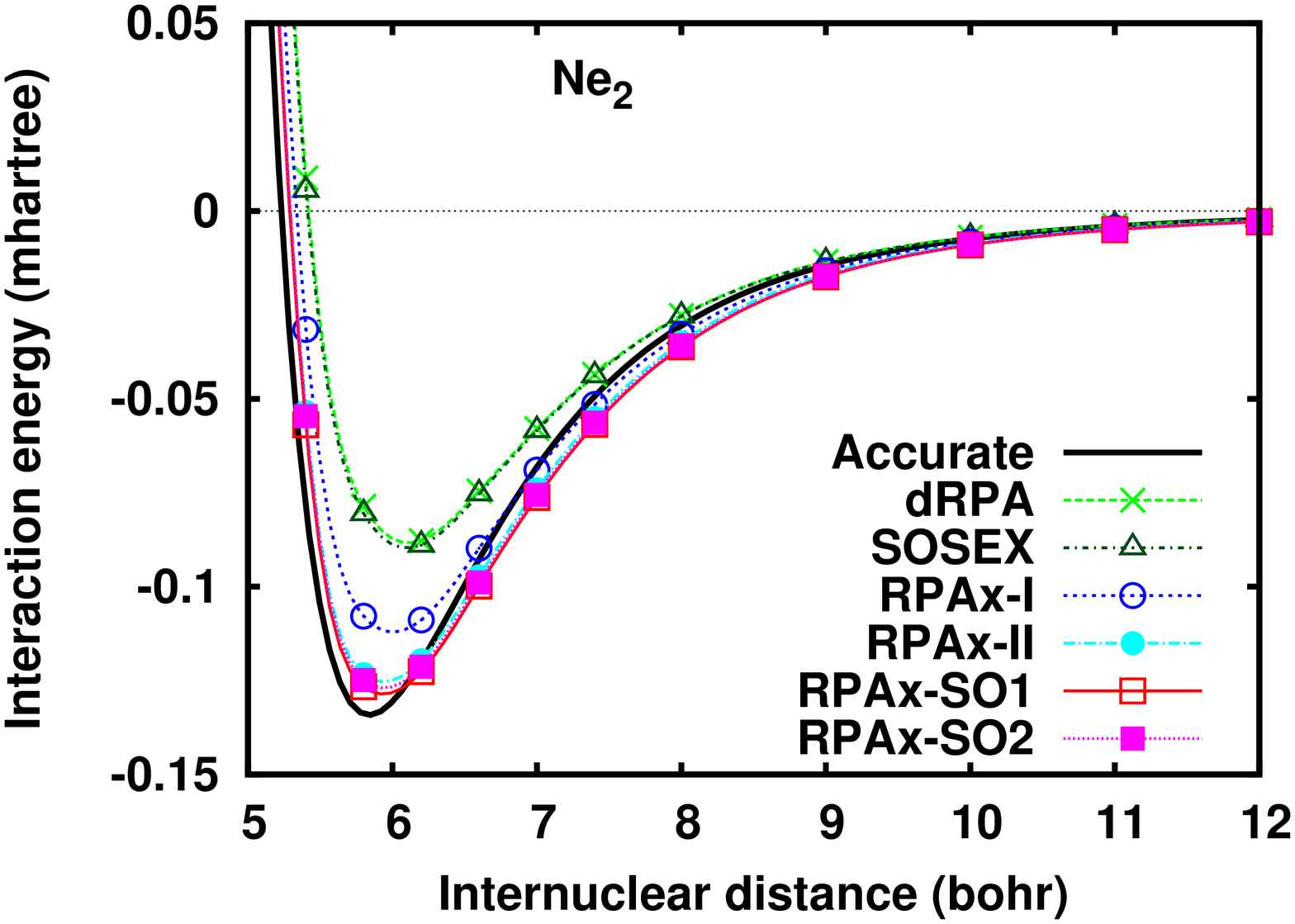}
\includegraphics[scale=0.30,angle=-90]{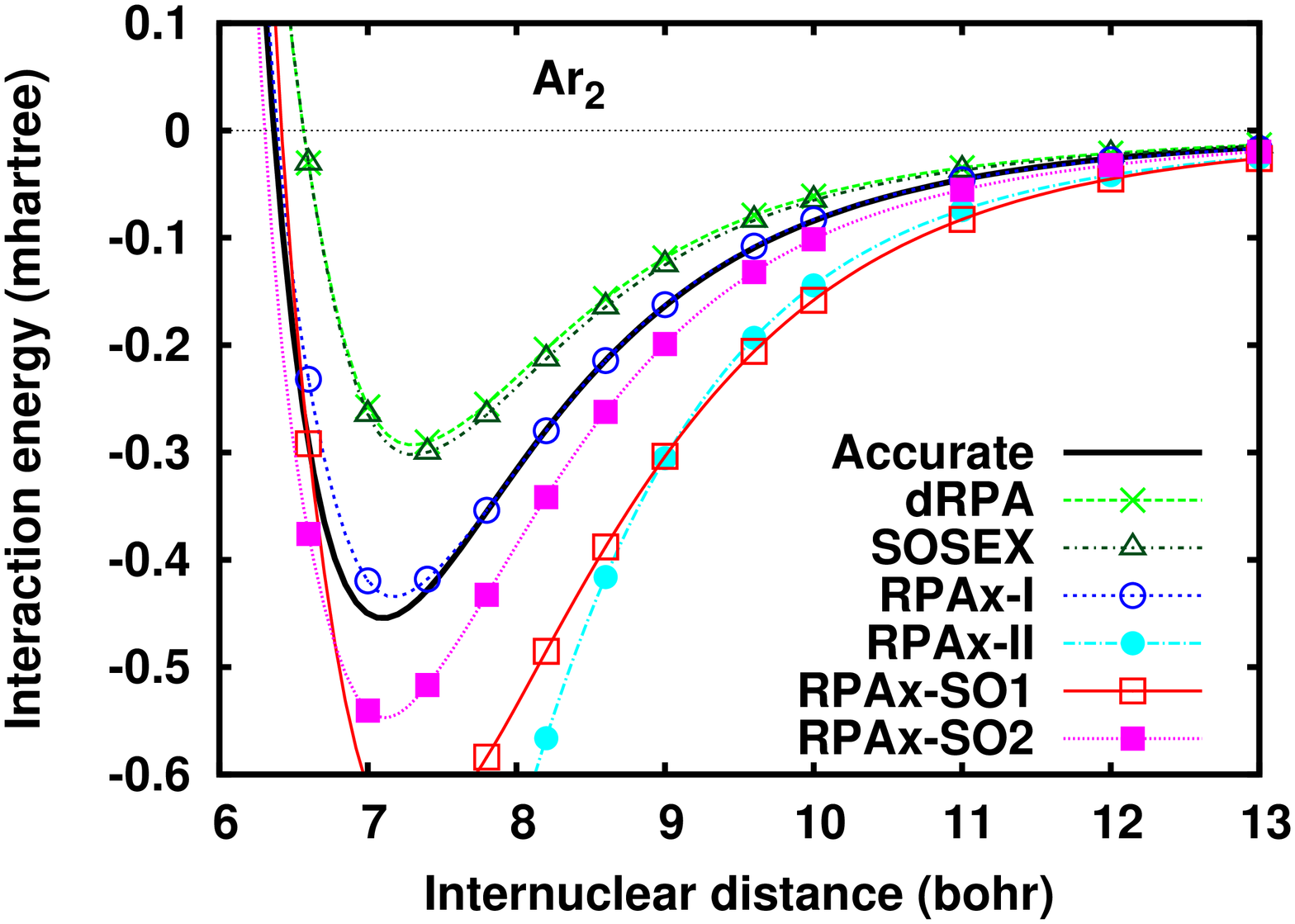}
\includegraphics[scale=0.30,angle=-90]{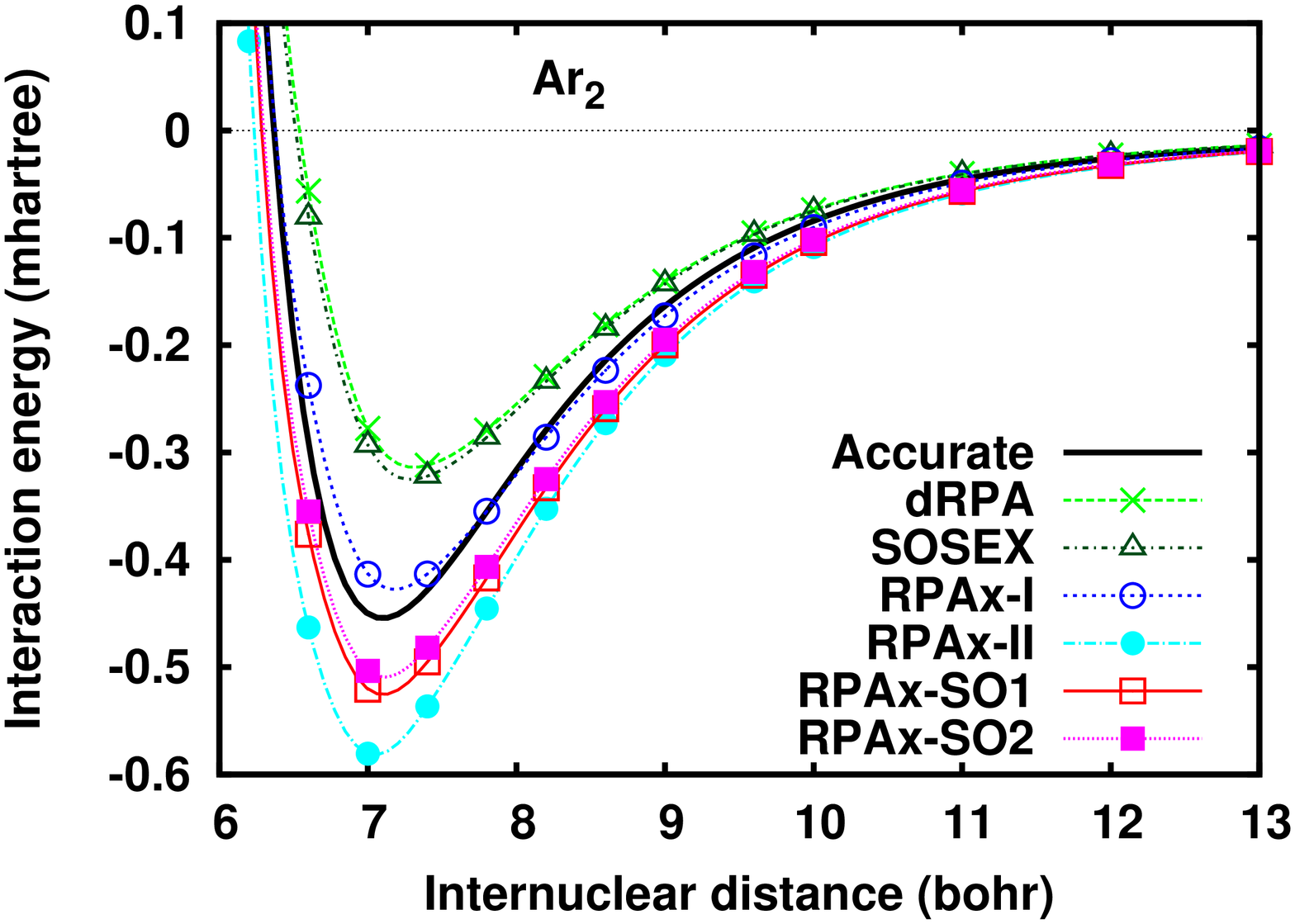}
\caption{(Color online) Interaction energy curves of He$_2$, Ne$_2$, and Ar$_2$ calculated by the full-range (left) and range-separated (right) RPA methods with the aug-cc-pV6Z basis set. Cubic splines are used to interpolate between the calculated points. The accurate curves are from Ref.~\onlinecite{TanToe-JCP-03}.
}
\label{fig:raregas}
\end{figure*}

Similarly, in the ring approximation with exchange terms, that we will refer to as RPAx (also sometimes referred to as RPA or time-dependent Hartree-Fock), the singlet and triplet amplitudes ${^1}\b{T}^{\lr}_{\RPAx}$ and ${^3}\b{T}^{\lr}_{\RPAx}$ are obtained by the equations
\begin{equation}
{^1}\b{B}^{\lr} + {^1}\b{A}^{\lr}  \, {^1}\b{T}^{\lr}_{\RPAx} + {^1}\b{T}^{\lr}_{\RPAx} {^1}\b{A}^{\lr} + {^1}\b{T}^{\lr}_{\RPAx} \, {^1}\b{B}^{\lr} \, {^1}\b{T}^{\lr}_{\RPAx} = \b{0},
\label{1TRPAx}
\end{equation}
and
\begin{equation}
{^3}\b{B}^{\lr} + {^3}\b{A}^{\lr}  \, {^3}\b{T}^{\lr}_{\RPAx} + {^3}\b{T}^{\lr}_{\RPAx} {^3}\b{A}^{\lr} + {^3}\b{T}^{\lr}_{\RPAx} \, {^3}\b{B}^{\lr} \, {^3}\b{T}^{\lr}_{\RPAx} = \b{0},
\label{3TRPAx}
\end{equation}
where ${^1}A^{\lr}_{ia,jb}=\Delta \epsilon_{ia,jb}+2\langle ib |aj \rangle^\lr-\langle ib |ja \rangle^\lr$, ${^3}A^{\lr}_{ia,jb}=\Delta \epsilon_{ia,jb}-\langle ib |ja \rangle^\lr$, and ${^3}B^{\lr}_{ia,jb}=-\langle ab |ji \rangle^\lr$. Using these amplitudes in the CCD correlation energy expression of Eq.~(\ref{EcCCDst}) gives what we call the RPAx-II long-range correlation energy (see, also, Refs.~\onlinecite{SzaOst-IJQC-77,SzaOst-JCP-77,Odd-AQC-78,AngLiuTouJan-JJJ-XX})
\begin{equation}
E_{c,\RPAxII}^{\lr} = \frac{1}{4} \tr \left[ {^1}\b{B}^{\lr} \, {^1}\b{T}^{\lr}_{\RPAx} + 3 {^3}\b{B}^{\lr} \, {^3}\b{T}^{\lr}_{\RPAx} \right],
\label{EcRPAxIIlr}
\end{equation}
which is equivalent to the plasmon formula expression of McLachlan and Ball~\cite{MclBal-RMP-64}. Using the same amplitudes in the alternative CCD correlation energy expression of Eq.~(\ref{EcCCDs}) gives another RPAx correlation energy which is the second approximation proposed by Szabo and Ostlund [Eq.~(3.22) of Ref.~\onlinecite{SzaOst-JCP-77}] as a zeroth iteration of the self-consistent RPA scheme~\cite{ShiMck-PRA-70,ShiMck-JCP-71,OstKar-CPL-71}
\begin{equation}
E_{c,\RPAxSOt}^{\lr} = \frac{1}{2} \tr \left[ {^1}\b{K}^{\lr} \, {^1}\b{T}^{\lr}_{\RPAx} \right].
\label{EcRPAxSOtlr}
\end{equation}
Equations~(\ref{EcRPAxIIlr}) and~(\ref{EcRPAxSOtlr}) are not equivalent because the ring approximation does not preserve the antisymmetry of the amplitudes with respect to the exchange of two spin-orbital indices. Using the same amplitudes in place of the singlet and triplet restricted amplitudes in the CCD correlation energy expression of Eq.~(\ref{EcCCDrst}) gives another alternative RPAx correlation energy corresponding to the first approximation proposed by Szabo and Ostlund [Eq.~(3.20) of Ref.~\onlinecite{SzaOst-JCP-77}, or Eq.~(17) of Ref.~\onlinecite{SzaOst-IJQC-77}] which is an alternative zeroth iteration of the self-consistent RPA scheme
\begin{equation}
E_{c,\RPAxSOo}^{\lr} = \frac{1}{2} \tr \left[ {^1}\b{B}^{\lr} \, \left( {^1}\b{T}^{\lr}_{\RPAx} - {^3}\b{T}^{\lr}_{\RPAx} \right) \right].
\label{EcRPAxSOolr}
\end{equation}
This last variant is the one preferred by Szabo and Ostlund because in a supermolecule approach it consistently gives a dispersion coefficient $C_6$ identical to the one given by the Casimir-Polder formula applied with the RPAx polarizabilities of the fragments, which is not the case for the other variants RPAx-II and $\RPAxSOt$. On the other hand, among the three RPAx methods proposed here, $\RPAxSOt$ has the advantage of involving only singlet excitations and thus is not subject to triplet instabilities. The RPAx method of Refs.~\onlinecite{TouGerJanSavAng-PRL-09,TouZhuAngSav-PRA-10}, that we will rename RPAx-I here, is yet another alternative correlation energy expression that involves only singlet excitations, but for which, as far as we know, the numerical integration over the adiabatic connection cannot be avoided (although in practice a single-point quadrature works well~\cite{ZhuTouSavAng-JCP-10}). It can be shown that the SOSEX, RPAx-I, RPAx-II, RPAx-SO1 and RPAx-SO2 correlation energies all correctly reduce to the MP2 correlation energy at second order in the electron-electron interaction, but dRPA does not. Finally, we note that an another RPAx correlation energy variant first proposed by Fukuda {\it et al.}~\cite{FukIwaSaw-PR-64} and defined as $2E_{c,\RPAxII}-E_{c,\text{MP2}}$ has also been discussed in the literature~\cite{SzaOst-IJQC-77,SzaOst-JCP-77,MosJezSza-IJQC-93,Hes-JCP-11}. It obviously correctly reduces to the MP2 correlation energy at second order, but numerical experience~\cite{Hes-JCP-11} shows that this variant gives very inaccurate correlation energies.

\begin{table*}[t]
\caption{Interaction energies (in kcal/mol) for the complexes of the S22 set from the range-separated RPA methods with the aug-cc-pVDZ basis set. For comparison, range-separated CCD results (without the ring approximation) are also reported. The geometries of complexes are taken from Ref.~\onlinecite{JurSpoCerHob-PCCP-06} and the reference interaction energies in the rightmost column are taken as the CCSD(T)/CBS estimates of Ref.~\onlinecite{TakHohMalMarShe-JCP-10}. Mean errors (ME), mean absolute errors (MAE) and mean absolute percentage errors (MA\%E) are given.
}
\label{tab:s22}
\begin{tabular}{llrrrrrrrr} \hline \hline
No. & Complex                          &    dRPA     &       SOSEX   &   RPAx-I  & RPAx-II  &  RPAx-SO1  & RPAx-SO2  &   CCD      & Reference \\
\hline                                                                                                                              
\multicolumn{7}{l}{Hydrogen-bonded complexes (HB7)}\\                                                                               
1   & (NH$_3$)$_2$                     &     -2.87   &    -2.92      &   -3.07   &  -3.26   &   -3.20    &   -3.18   &  -3.20     &   -3.17 \\
2   & (H$_2$O)$_2$                     &     -5.16   &    -5.23      &   -5.33   &  -5.42   &   -5.40    &   -5.39   &  -5.41     &   -5.02 \\
3   & Formic acid dimer                &    -20.30   &   -20.55      &  -20.81   & -20.98   &  -20.94    &  -20.98   & -20.94     &  -18.80 \\
4   & Formamide dimer                  &    -16.51   &   -16.68      &  -17.03   & -17.48   &  -17.32    &  -17.27   & -17.25     &  -16.12 \\
5   & Uracil dimer $C_{2h}$            &    -21.03   &   -21.36      &  -21.80   & -22.58   &  -22.04    &  -22.15   & -22.00     &  -20.69 \\
6   & 2-pyridoxine/2-aminopyridine     &    -17.07   &   -17.28      &  -17.81   & -18.89   &  -18.25    &  -18.20   & -18.08     &  -17.00 \\
7   & Adenine/thymine WC               &    -16.53   &   -16.73      &  -17.29   & -18.25   &  -17.81    &  -17.69   & -17.62     &  -16.74 \\
    & ME                               &     -0.28   &    -0.46      &   -0.80   &  -1.33   &   -1.06    &   -1.05   &  -0.99     &    0.00 \\
    & MAE                              &      0.42   &     0.53      &    0.83   &   1.33   &    1.06    &    1.05   &   0.99     &    0.00 \\
    & MA\%E                            &      3.7\%  &     4.3\%     &    5.6\%  &   8.6\%  &    6.8\%   &    6.6\%  &   6.4\%    &    0.0\% \\
\\                                                                                                                                  
\multicolumn{7}{l}{Complexes with predominant dispersion contribution (WI8)}\\                                                      
8   & (CH$_4$)$_2$                     &     -0.30   &    -0.31      &   -0.42   &  -0.56   &    -0.53   &   -0.51   &   -0.51    &   -0.53 \\
9   & (C$_2$H$_4$)$_2$                 &     -0.97   &    -1.02      &   -1.28   &  -1.66   &    -1.52   &   -1.47   &   -1.45    &   -1.50 \\
10  & Benzene/CH$_4$                   &     -0.92   &    -0.98      &   -1.23   &  -1.75   &    -1.47   &   -1.43   &   -1.40    &   -1.45 \\
11  & Benzene dimer $C_{2h}$           &     -1.27   &    -1.38      &   -2.05   &  -4.28   &    -2.72   &   -2.61   &   -2.40    &   -2.62 \\
12  & Pyrazine dimer                   &     -2.99   &    -3.10      &   -3.78   &  -6.12   &    -4.49   &   -4.34   &   -4.14    &   -4.20 \\
13  & Uracil dimer $C_2$               &     -8.22   &    -8.46      &   -9.38   & -11.93   &   -10.25   &  -10.13   &   -9.94    &   -9.74 \\
14  & Indole/benzene                   &     -2.58   &    -2.75      &   -3.70   &  -7.12   &    -4.64   &   -4.48   &   -4.17    &   -4.59 \\
15  & Adenine/thymine stack            &     -9.38   &    -9.68      &  -10.97   & -15.14   &   -12.23   &  -12.02   &  -11.72    &  -11.66 \\
    & ME                               &      1.21   &     1.08      &    0.43   &  -1.53   &    -0.20   &   -0.09   &    0.07    &    0.00 \\
    & MAE                              &      1.21   &     1.08      &    0.43   &   1.53   &     0.20   &    0.13   &    0.13    &    0.00 \\
    & MA\%E                            &     34.3\%  &    31.2\%     &   13.9\%  &  31.7\%  &     3.1\%  &    2.5\%  &    3.8\%   &    0.0\% \\
\\                                                                                                                                  
\multicolumn{7}{l}{Mixed complexes (MI7)}\\                                                                                         
16  & Ethene/ethyne                    &     -1.31   &    -1.36      &  -1.48    &  -1.67   &     -1.58  &   -1.57   &    -1.55   &   -1.51 \\
17  & Benzene/H$_2$O                   &     -2.90   &    -2.96      &  -3.16    &  -3.52   &     -3.34  &   -3.30   &    -3.29   &   -3.29 \\
18  & Benzene/NH$_3$                   &     -1.83   &    -1.88      &  -2.11    &  -2.57   &     -2.33  &   -2.29   &    -2.27   &   -2.32 \\
19  & Benzene/HCN                      &     -4.20   &    -4.31      &  -4.54    &  -4.98   &     -4.72  &   -4.71   &    -4.65   &   -4.55 \\
20  & Benzene dimer $C_{2v}$           &     -1.92   &    -2.00      &  -2.39    &  -3.40   &     -2.77  &   -2.70   &    -2.61   &   -2.71 \\
21  & Indole/benzene T-shape           &     -4.54   &    -4.65      &  -5.17    &  -6.57   &     -5.66  &   -5.57   &    -5.44   &   -5.62 \\
22  & Phenol dimer                     &     -6.48   &    -6.62      &  -7.07    &  -8.16   &     -7.49  &   -7.43   &    -7.35   &   -7.09 \\
    & ME                               &      0.56   &     0.47      &   0.17    &  -0.54   &     -0.11  &   -0.07   &    -0.01   &    0.00 \\
    & MAE                              &      0.56   &     0.47      &   0.17    &   0.54   &      0.11  &    0.09   &     0.11   &    0.00 \\
    & MA\%E                            &     15.8\%  &    13.5\%     &   5.0\%   &  13.6\%  &      2.7\% &    2.2\%  &     2.6\%   &    0.0\% \\  
\\                                                                                                                                  
    & total ME                         &      0.53   &     0.40      &  -0.04    &  -1.15   &     -0.44  &   -0.39   &     -0.30  &    0.00  \\
    & total MAE                        &      0.75   &     0.71      &   0.47    &   1.15   &      0.44  &    0.41   &      0.40  &    0.00 \\
    & total MA\%E                      &     18.7\%  &    17.0\%     &   8.4\%   & 18.6\%  &      4.1\% &    3.7\%  &       4.3\% &  0.0\% \\
\hline\hline
\end{tabular}
\end{table*}

\section{Computational details}

All calculations have been done with a development version of MOLPRO 2008~\cite{Molproshort-PROG-08}, implementing equations~(\ref{eq1TdRPA})-(\ref{EcRPAxSOolr}). We first perform a self-consistent RSH calculation with the short-range Perdew-Burke-Ernzerhof (PBE) exchange-correlation functional of Ref.~\onlinecite{GolWerStoLeiGorSav-CP-06} and add the long-range RPA correlation energies calculated with RSH orbitals. The range separation parameter is taken at $\mu=0.5$ bohr$^{-1}$, according to previous studies~\cite{GerAng-CPL-05a}, without trying to readjust it. For the rare-gas dimers, we also carry out full-range RPA calculations using PBE orbitals~\cite{PerBurErn-PRL-96} for comparison. The Riccati equation~(\ref{eq1TdRPA}) is solved by decomposing the matrix $^1\b{L}^\lr$ into diagonal and off-diagonal parts and iteratively extracting $^1\b{T}^\lr_\dRPA$ from its product with the diagonal part and updating it in the other terms, and similarly for Eqs.~(\ref{1TRPAx}) and~(\ref{3TRPAx}). For RPAx-I calculations, the adiabatic-connection integration is performed by a 8-point Gauss-Legendre quadrature for the rare-gas dimers, and by a single-point quadrature [Eq.~(14) of Ref.~\onlinecite{ZhuTouSavAng-JCP-10}] for the S22 set. We use the correlation-consistent basis sets of Dunning~\cite{Dun-JCP-89,WilMouDun-JMS-96}. Core electrons are kept frozen (i.e. only excitations of valence electrons are considered). Basis set superposition error (BSSE) is removed by the counterpoise method. The geometries of the complexes of the S22 set are taken from Ref.~\onlinecite{JurSpoCerHob-PCCP-06}. The geometries of the isolated monomers are fixed to those in the dimers, thus the so-called monomer deformation energy is not included in the interaction energy. For each method, mean error (ME), mean absolute error (MAE) and mean absolute percentage error (MA\%E) are given using as a reference the CCSD(T) values extrapolated to the complete basis set (CBS) limit of Takatani {\it et al.}~\cite{TakHohMalMarShe-JCP-10}.

In our present, most basic implementation, the computational cost of all the RPA methods used here formally scales as $N_v^3 N_o^3$ for large basis sets, where $N_v$ and $N_o$ are the numbers of virtual and occupied orbitals, respectively. The computational cost of the CCD (or CCSD) method without the ring approximation is higher and it scales as $N_v^4 N_o^2$ for large basis sets~\cite{LeeRic-CPL-88}. Of course, far better scalings should be obtained by using integral-direct methods and resolution-of-identity/Cholesky-decomposition techniques~\cite{ScuHenSor-JCP-08}.

\section{Results}

The interaction energy curves of He$_2$, Ne$_2$, and Ar$_2$ calculated by the full-range and range-separated RPA methods are compared in Fig.~\ref{fig:raregas}. We use the large aug-cc-pV6Z basis set to ensure that the full-range calculations are converged. Full-range dRPA and SOSEX strongly underestimate the interaction energies, while full-range RPAx-II and RPAx-SO1 strongly overestimate them. The best full-range methods are RPAx-I and RPAx-SO2, which is in agreement with the recent study of He{\ss}elmann~\cite{Hes-JCP-11,TouZhuSavJanAng-JJJ-XX-note}. In passing, we note that the full-range RPAx-I method better performs for Ne$_2$ and Ar$_2$ when using PBE orbitals than when using HF orbitals, as done in Ref.~\onlinecite{TouZhuAngSav-PRA-10}. Range separation greatly improves the accuracy of all the RPA variants. However, range-separated dRPA and SOSEX still underestimate the interaction energies, and range-separated RPAx-II significantly overestimates the interaction energy of Ar$_2$. Range-separated RPAx-I, RPAx-SO1, and RPAx-SO2 give the most reasonable interaction energy curves.

The interaction energies for the complexes of the S22 set calculated with the range-separated RPA methods with the aug-cc-pVDZ basis set are given in Table~\ref{tab:s22}. For comparison, range-separated CCD results (without the ring approximation) are also reported. Although the aug-cc-pVDZ basis set may appear small, range-separated RPA methods are weakly dependent on the basis size~\cite{TouGerJanSavAng-PRL-09,TouZhuAngSav-PRA-10}, and indeed it was estimated in Ref.~\onlinecite{ZhuTouSavAng-JCP-10} that when going from the aug-cc-pVDZ to the aug-cc-pVTZ basis set the range-separated RPAx-I interaction energies of the S22 set are lower by at most 7\%, and the corresponding total MA\%E decreases by less than 2\%. Therefore, we believe that the aug-cc-pVDZ basis set is sufficient to compare the different range-separated RPA methods.

The S22 set includes seven hydrogen-bonded complexes (HB7 subset͒), eight weakly interacting complexes with predominant dispersion contributions (WI8 subset͒), and seven mixed complexes featuring also multipole interactions (MI7 subset). The trends are quite different for the HB7 subset͒ on the one hand, and the WI8 and MI7 subsets on the other hand. It was previously argued that the general overestimation of the interaction energies of hydrogen-bonded complexes is due to the approximate short-range density functional~\cite{GolLeiManMitWerSto-PCCP-08,ZhuTouSavAng-JCP-10}. The fact that dRPA and SOSEX give the smallest MAEs for the HB7 subset͒ is thus not believed to be significant but rather due to a compensation of errors between an underestimated long-range contribution and an overestimated short-range contribution. This is corroborated by the relatively large overestimation of the interaction energies of this subset by range-separated CCD which should most accurately describe the long-range correlation energies. We will thus focus our analysis on the WI8 and MI7 subsets.

For the WI8 and MI7 subsets, dRPA gives largely underestimated interaction energies, with MA\%Es of 34.3\% and 15.8\%, respectively. SOSEX barely improves dRPA with MA\%Es of 31.2\% and 13.5\%, respectively. This may not be surprising since, in the limit of large separation, SOSEX only adds exponentially decaying exchange interactions between the monomers, but does not change the coupled-cluster amplitudes and thus does not change the polarizabilities of the monomers. The RPAx-I method of Refs.~\onlinecite{TouGerJanSavAng-PRL-09,TouZhuAngSav-PRA-10}, which incorporates exchange effects in the monomers, greatly reduces the underestimation of the interaction energies, with MA\%Es of 13.9\% and 5.0\%, respectively. The RPAx-II variant, which may be seen as the most straightforward way of defining a closed-shell ring CCD with exchange terms, is disappointingly inaccurate. It overestimates the interaction energies by about the same amount that dRPA underestimates them. Finally, the two variants RPAx-SO1 and RPAx-SO2 give remarkably accurate interaction energies, with MA\%Es of 3.1\% and 2.7\% for RPAx-SO1, and 2.5\% and 2.2\% for RPAx-SO2. They are globally as accurate as range-separated CCD without the ring approximation. However, it must be noted that RPAx-SO1 and RPAx-SO2 tend to overestimate dispersion energies, while RPAx-I underestimates them. Therefore, increasing the basis size will likely increase the MA\%Es of RPAx-SO1 and RPAx-SO2, while it will decrease the MA\%E of RPAx-I.

\section{Conclusion}

We have studied various RPA variants that can be cast in the form of closed-shell ring CCD approximations. We have tested these variants with range separation, i.e. by combining a long-range RPA-type approximation with a short-range density-functional approximation, on rare-gas dimers and on the weakly interacting complexes of the S22 set. Among all these variants, the ones first proposed by Szabo and Ostlund~\cite{SzaOst-IJQC-77,SzaOst-JCP-77}, called here RPAx-SO1 [Eq.~(\ref{EcRPAxSOolr})] and RPAx-SO2 [Eq.~(\ref{EcRPAxSOtlr})], give the most accurate dispersion energies. The other variants tend to either strongly underestimate (dRPA and SOSEX) or strongly overestimate (RPAx-II) the interaction energies. For comparison, we have also reported results from the RPAx-I method of Refs.~\onlinecite{TouGerJanSavAng-PRL-09,TouZhuAngSav-PRA-10}, which is not based on a ring CCD approximation but on the adiabatic connection formula, and which gives reasonable interaction energies as well. From a practical point of view, RPAx-SO2 appears to be the most convenient variant since, contrary to RPAx-I, it does not use any numerical adiabatic-connection integration and, contrary to RPAx-SO1, it involves only singlet excitations and is thus not subject to triplet instabilities.

\section*{Acknowledgments}
This work was supported by ANR (French national research agency) via contract number ANR-07-BLAN-0272 (Wademecom). We thank D. Mukherjee (Kolkata, India) for discussions.

\appendix
\section{CCD correlation energy}
\label{app:CCD}
In this appendix, we review several equivalent CCD correlation energy expressions, in view of justifying the different ring CCD variants.

\subsection{CCD correlation energy in spin-orbital basis}
The spin-unrestricted CCD wave function {\it ansatz} is
\begin{equation}
\ket{\Psi_\CCD} = \exp \left( \hat{T}_2 \right) \ket{\Phi},
\end{equation}
where $\ket{\Phi}$ is a single-determinant reference wave function, and $\hat{T}_2$ is the cluster operator for double excitations which is written in a spin-orbital basis as
\begin{equation}
\hat{T}_2 = \frac{1}{4} \sum_{ijab} t_{ij}^{ab} \hat{a}_a^\dag \hat{a}_i \hat{a}_b^\dag \hat{a}_j,
\end{equation}
where $i,j$ and $a,b$ refer to occupied and virtual spin-orbitals, respectively, and the amplitudes $t_{ij}^{ab}$ must be antisymmetric with respect to any exchange of two indices: $t_{ij}^{ab}=-t_{ji}^{ab}=-t_{ij}^{ba}=t_{ji}^{ba}$. The CCD correlation energy is obtained by the transition formula
\begin{eqnarray}
E_c^\CCD &=& \bra{\Phi} \hat{H} \ket{\Psi_\CCD} - \bra{\Phi} \hat{H} \ket{\Phi} = \bra{\Phi} \hat{H} \hat{T}_2 \ket{\Phi}
\nonumber\\
&=& \frac{1}{4} \sum_{ijab} \langle ab ||ij \rangle  t_{ij}^{ab} = \frac{1}{4} \tr \left[ \b{B} \b{T} \right],
\end{eqnarray}
where $B_{ia,jb}=\langle ab ||ij \rangle$ are the antisymmetrized two-electron integrals over real spin orbitals and $T_{ia,jb}=t_{ij}^{ab}$ is the amplitude matrix. Using the antisymmetry of the amplitudes, the CCD correlation energy can also be written as
\begin{eqnarray}
E_c^\CCD &=& \frac{1}{2} \sum_{ijab} \langle ab |ij \rangle  t_{ij}^{ab} = \frac{1}{2} \tr \left[ \b{K} \b{T} \right],
\end{eqnarray}
where $K_{ia,jb}=\langle ab |ij \rangle$ are the two-electron integrals.

\subsection{CCD correlation energy in spatial-orbital basis for closed-shell systems}

\subsubsection{Expression in terms of the singlet and triplet amplitudes}

For spin-restricted closed-shell calculations, all the matrices in the spin-orbital excitation basis encountered so far (e.g., $\b{A}$, $\b{B}$, $\b{K}$, $\b{T}$) have the following spin block structure
\begin{equation}
\b{C} = \left( 
  \begin{array}{cccc} 
   \b{C}_{\ua \ua,\ua \ua} & 
   \b{C}_{\ua \ua,\da \da} & 
   \b{0}                       & \b{0}                      \\ 
   \b{C}_{\da \da,\ua \ua} & 
   \b{C}_{\da \da,\da \da} & 
   \b{0}                       & \b{0}                      \\
   \b{0}                       & \b{0}                       &  
   \b{C}_{\ua \da,\ua \da} & \b{C}_{\ua \da,\da \ua}\\ 
   \b{0}                       & \b{0}                       & 
   \b{C}_{\da \ua,\ua \da} & \b{C}_{\da \ua,\da \ua}\\ 
  \end{array} 
  \right),
\end{equation}
and can be brought to a block-diagonal spin-adapted matrix $\b{\tilde{C}} = \b{U}^{\T} \, \b{C} \, \b{U}$ by the orthogonal transformation
\begin{equation}
\b{U} = \frac{1}{\sqrt{2}} \left( 
   \begin{array}{cccc} 
   \b{1} &  \b{1}  & \b{0} &  \b{0} \\ 
   \b{1} & -\b{1}  & \b{0} &  \b{0} \\
   \b{0} &  \b{0}  & \b{1} &  \b{1} \\
   \b{0} &  \b{0}  & \b{1} & -\b{1} \\
   \end{array} 
   \right). 
\end{equation}
Applying this transformation to the matrix $\b{B}$ gives the following decomposition into singlet and triplet excitations
\begin{equation}
\b{\tilde{B}} =  \left( 
   \begin{array}{cccc} 
   ^1\b{B} & \b{0} & \b{0} & \b{0} \\
     \b{0} & ^3\b{B} & \b{0} & \b{0} \\
     \b{0} & \b{0} & ^3\b{B} & \b{0} \\
     \b{0} & \b{0} & \b{0} & -{^3}\b{B}\\
   \end{array} 
   \right), 
\end{equation}
where ${^1}B_{ia,jb}=2\braket{ab}{ij}-\braket{ab}{ji}$ and ${^3}B_{ia,jb}=-\braket{ab}{ji}$, with $i,j$ referring now to occupied spatial orbitals and $a,b$ to virtual spatial orbitals. Notice the minus sign for the last triplet block. Using Kramers symmetry for spin-conserving real coupled-cluster amplitudes (see, e.g., Ref.~\onlinecite{WanGauWul-JCP-08}), one can show that spin adaptation of the matrix $\b{T}$ leads to a similar form
\begin{equation}
\b{\tilde{T}} =  \left( 
   \begin{array}{cccc} 
   ^1\b{T} & \b{0} & \b{0} & \b{0} \\
     \b{0} & ^3\b{T} & \b{0} & \b{0} \\
     \b{0} & \b{0} & ^3\b{T} & \b{0} \\
     \b{0} & \b{0} & \b{0} & -{^3}\b{T}\\
   \end{array} 
   \right), 
\end{equation}
where $^1T_{ia,jb}=T_{i\ua a\ua, j\ua b\ua}+T_{i\ua a\ua, j\da b\da}$ and $^3T_{ia,jb}=T_{i\ua a\ua, j\ua b\ua}-T_{i\ua a\ua, j\da b\da}=T_{i\ua a\da, j\da b\ua}$. The CCD correlation energy can thus be expressed as
\begin{equation}
E_{c}^{\CCD} = \frac{1}{4} \tr \left[ {^1}\b{B} \, {^1}\b{T} + 3 {^3}\b{B} \, {^3}\b{T} \right].
\label{EcCCDst}
\end{equation}
Spin adaptation of the matrix $\b{K}$ gives only a contribution from the singlet excitations
\begin{equation}
\b{\tilde{K}} =  \left( 
   \begin{array}{cccc} 
   ^1\b{K} & \b{0} & \b{0} & \b{0} \\
     \b{0} & \b{0} & \b{0} & \b{0} \\
     \b{0} & \b{0} & \b{0} & \b{0} \\
     \b{0} & \b{0} & \b{0} & \b{0}\\
   \end{array} 
   \right), 
\end{equation}
where ${^1}K_{ia,jb}=2\braket{ab}{ij}$, which leads to an alternative form for the CCD correlation energy
\begin{equation}
E_{c}^{\CCD} = \frac{1}{2} \tr \left[ {^1}\b{K} \, {^1}\b{T} \right].
\label{EcCCDs}
\end{equation}

\subsubsection{Expression in terms of the restricted amplitudes}

In practice, the CCD correlation energy is normally calculated starting from the spin-restricted closed-shell CCD wave function {\it ansatz}
\begin{equation}
\ket{\Psi_\CCD} = \exp \left( ^\R\hat{T}_2 \right) \ket{\Phi},
\end{equation}
where the restricted cluster operator $^\R\hat{T}_2$ is written in a spatial-orbital basis as
\begin{equation}
^\R\hat{T}_2 = \frac{1}{2} \sum_{ijab} {^\R}t_{ij}^{ab} \hat{E}_{ai} \hat{E}_{bj},
\label{RT2}
\end{equation}
where $\hat{E}_{ai}=\hat{a}_{a\uparrow}^\dag \hat{a}_{i\uparrow} + \hat{a}_{a\downarrow}^\dag \hat{a}_{i\downarrow}$ is the singlet excitation operator and ${^\R}t_{ij}^{ab}$ are the restricted amplitudes which must be symmetric with respect to the exchange of both $i,j$ and $a,b$, i.e., ${^\R}t_{ij}^{ab}={^\R}t_{ji}^{ba}$, but not antisymmetric with respect to the exchange of only two indices. The CCD correlation energy is obtained by the transition formula
\begin{eqnarray}
E_c^\CCD &=& \bra{\Phi} \hat{H} \ket{\Psi_\CCD} - \bra{\Phi} \hat{H} \ket{\Phi} = \bra{\Phi} \hat{H} \, {^\R}\hat{T}_2 \ket{\Phi}
\nonumber\\
&=& \sum_{ijab} \left( 2 \langle ab |ij \rangle  - \langle ab |ji \rangle \right) {^\R}t_{ij}^{ab} = \tr \left[ {^1}\b{B} \, {^\R}\b{T} \right],
\label{EcCCD}
\end{eqnarray}
where ${^\R}T_{ia,jb}={^\R}t_{ij}^{ab}$.

\subsubsection{Expression in terms of the singlet and triplet restricted amplitudes}

Another equivalent correlation energy expression can be obtained by decomposing the restricted amplitudes into spin-singlet and spin-triplet components. Indeed, the restricted cluster operator can be decomposed as (see, e.g., Ref.~\onlinecite{HelJorOls-BOOK-02,Pal-JCP-77})
\begin{equation}
^\R\hat{T}_2 = \frac{1}{2} \sum_{ijab} \left( {^{1,\R}}t_{ij}^{ab} \hat{S}_{aibj} + {^{3,\R}}t_{ij}^{ab} \hat{T}_{aibj} \right),
\label{T2st}
\end{equation}
where ${^{1,\R}}t_{ij}^{ab}$ are singlet restricted amplitudes
\begin{equation}
{^{1,\R}}t_{ij}^{ab} = {^\R}t_{ji}^{ab} + {^\R}t_{ij}^{ab},
\label{1tijab}
\end{equation}
which are totally symmetric (i.e., ${^{1,\R}}t_{ij}^{ab}={^{1,\R}}t_{ji}^{ab}= {^{1,\R}}t_{ij}^{ba} ={^{1,\R}}t_{ji}^{ba}$), and ${^{3,\R}}t_{ij}^{ab}$ are the triplet restricted amplitudes
\begin{equation}
{^{3,\R}}t_{ij}^{ab} = {^\R}t_{ji}^{ab} - {^\R}t_{ij}^{ab},
\label{3tijab}
\end{equation}
which are totally antisymmetric (i.e., ${^{3,\R}}t_{ij}^{ab}=-{^{3,\R}}t_{ji}^{ab}=-{^{3,\R}}t_{ij}^{ba} ={^{3,\R}}t_{ji}^{ba}$). In Eq.~(\ref{T2st}), $\hat{S}_{aibj}$ is the singlet double-excitation operator
\begin{equation}
\hat{S}_{aibj} = \hat{S}_{ai}^{0,0} \hat{S}_{bj}^{0,0} = \frac{1}{2} \hat{E}_{ai} \hat{E}_{bj},
\end{equation}
constructed with the singlet single-excitation operator 
\begin{equation}
\hat{S}_{ai}^{0,0} = \frac{1}{\sqrt{2}} \left( \hat{a}_{a\uparrow}^\dag \hat{a}_{i\uparrow} + \hat{a}_{a\downarrow}^\dag \hat{a}_{i\downarrow} \right) = \frac{1}{\sqrt{2}} \hat{E}_{ai},
\end{equation}
and $\hat{T}_{aibj}$ is the triplet double-excitation operator
\begin{equation}
\hat{T}_{aibj} = \hat{T}_{ai}^{1,1} \hat{T}_{bj}^{1,-1} - \hat{T}_{ai}^{1,0} \hat{T}_{bj}^{1,0} + \hat{T}_{ai}^{1,-1} \hat{T}_{bj}^{1,1} =\hat{E}_{aj}\hat{E}_{bi} + \frac{1}{2} \hat{E}_{ai}\hat{E}_{bj},
\end{equation}
constructed with the triplet single-excitation operators
\begin{subequations}
\begin{equation}
\hat{T}_{ai}^{1,1} = - \hat{a}_{a\uparrow}^\dag \hat{a}_{i\downarrow},
\end{equation}
\begin{equation}
\hat{T}_{ai}^{1,0} = \frac{1}{\sqrt{2}} \left( \hat{a}_{a\uparrow}^\dag \hat{a}_{i\uparrow} - \hat{a}_{a\downarrow}^\dag \hat{a}_{i\downarrow} \right),
\end{equation}
\begin{equation}
\hat{T}_{ai}^{1,-1} = \hat{a}_{a\downarrow}^\dag \hat{a}_{i\uparrow}.
\end{equation}
\end{subequations}
Using the symmetry properties of ${^{1,\R}}t_{ij}^{ab}$ and ${^{3,\R}}t_{ij}^{ab}$, it is easy to check that Eqs.~(\ref{RT2}) and~(\ref{T2st}) are equivalent. Combining Eqs.~(\ref{1tijab}) and~(\ref{3tijab}) leads to the decomposition of the restricted amplitudes into spin components
\begin{equation}
{^\R}t_{ij}^{ab} = \frac{1}{2} \left( {^{1,\R}}t_{ij}^{ab} - {^{3,\R}}t_{ij}^{ab} \right),
\end{equation}
and the CCD correlation energy [Eq.~(\ref{EcCCD})] can thus be written as
\begin{eqnarray}
E_c^\CCD = \frac{1}{2} \tr \left[ {^1}\b{B} \left( {^{1,\R}}\b{T} - {^{3,\R}}\b{T} \right) \right].
\label{EcCCDrst}
\end{eqnarray}
This corresponds to the definition of singlet and triplet contributions to the correlation energy, $E_c^\S=(1/2) \tr [ {^1}\b{B} \, {^{1,\R}}\b{T} ]$ and $E_c^\T=-(1/2) \tr [ {^1}\b{B} \, {^{3,\R}}\b{T} ]$. By using the symmetry properties of ${^{1,\R}}\b{T}$ and ${^{3,\R}}\b{T}$, one can show that they are equivalent to the more usual expressions in terms of the restricted amplitudes (see, e.g., Ref.~\onlinecite{Klo-MP-01}): $E_c^\S=(1/4) \tr [ ( {^1}\b{B} - 3 {^3}\b{B} ) {^\R}\b{T} ]$ and $E_c^\T=(3/4) \tr [ ({^1}\b{B} + {^3}\b{B} ) {^\R}\b{T} ]$.


\end{document}